**Distributed feedback terahertz frequency quantum cascade lasers with dual periodicity gratings**

F. Castellano,[1] S. Zanotto,[1] L. H. Li,[2] A. Pitanti,[1] A. Tredicucci,[3] E. H. Linfield,[2] A. G. Davies,[2] and M. S. Vitiello[1]

[1]*NEST, CNR - Istituto Nanoscienze and Scuola Normale Superiore, Piazza San Silvestro 12, 56127, Pisa, Italy*
[2]*School of Electronic and Electrical Engineering, University of Leeds, Leeds LS2 9JT, UK*
[3] *Dipartimento di Fisica, Università di Pisa, Largo Pontecorvo 3 and NEST, CNR - Istituto Nanoscienze, 56127 Pisa, Italy*

We have developed terahertz frequency quantum cascade lasers that exploit a double-periodicity distributed feedback grating to control the emission frequency and the output beam direction independently. The spatial refractive index modulation of the gratings necessary to provide optical feedback at a fixed frequency and, simultaneously, a far-field emission pattern centered at controlled angles, was designed through use of an appropriate wavevector scattering model. Single mode THz emission at angles tuned by design between 0° and 50° was realized, leading to an original phase-matching approach, lithographically independent, for highly collimated THz QCLs.

Terahertz (THz) frequency quantum cascade lasers (QCLs) have undergone rapid development in performance since their first demonstration, [1] finding potential application in a number of fields including astronomy, security screening, biomedicine, and cultural heritage, *inter alia* [2,3]. Operating across the 1.2–5 THz range, QCLs can provide high peak output powers (1 W) [4], a high spectral purity [5,6], frequency, phase and amplitude stability [7–9], and an ultra-broadband gain spanning an octave in frequency [10,11] at temperatures ≤199 K [12].

Most of the mentioned applications require sources with a low divergent spatial profile in the far-field as well as a fine spectral control of the emitted radiation. However, the double-metal waveguides conventionally employed to maximize the THz QCL operating temperature [12] suffer from a lack of efficient extraction and a poor collimation of the output radiation [13], owing to the sub-wavelength dimensions of the resonant cavities. Also, the strong longitudinal confinement provided by such microstrip waveguide configuration typically induces the laser to operate in a multimode regime.

Distributed feedback (DFB) is commonly employed in semiconductor lasers to tailor the shape, symmetry, and frequency of the optical resonator eigenmodes, while simultaneously allowing stable single-mode operation. A DFB operates by introducing a periodic refractive index modulation along the propagation direction, which provides scattering between two guided states (feedback) or between a guided state and a radiative one (extraction), so that distributed optical feedback is achieved by coupling two counter-propagating modes through one of the spatial harmonics of a refractive index modulation. In fact, the latter is never purely sinusoidal and a grating with periodicity *L* contains spatial harmonics with wavevector $k_n = n/L$. Depending on whether the first, second or third harmonic is used for feedback ($n = 1, 2, 3$), the grating is described to be of the first, second or third order, respectively [14]. In the case of second [15, 16] and third order [17] gratings, although a high spatial harmonic is used for optical feedback ($k_{fb} = k_2, k_3...$), the fundamental frequency is employed for radiation extraction ($k_e = k_1$). The ratio between the feedback wavevector $k_{fb}$ and the extraction wavevector $k_e$ determines the behavior of the laser: second order gratings emit perpendicularly to the waveguide plane, while third order gratings emit parallel to the waveguide plane only if the effective mode index n=3.

The difficulty of achieving controlled spectral and spatial beam patterns from THz QCLs has recently been addressed by engineering two-dimensional photonic crystal lasers [18] and edge-emitting third-order (periodic) DFBs [17,19]. Despite the clear advantages in terms of emission profiles, there are some challenges associated with the third-order DFBs: a perfect phase-matching condition must be established lithographically and, being highly dependent on the lasing frequency, it requires high precision, commonly a demanding task. In all of the examples discussed above, different spatial harmonics of periodic gratings were used to provide optical-feedback and power extraction. Recently, however, non-periodic feedback gratings have been engineered and investigated in an attempt to overcome some of the inherent limitations of standard DFBs, in terms of output beam quality and optical power [20–23].

In this letter, we report on the development of THz QCLs employing feedback gratings with a double-periodicity. By exploiting two unrelated spatial frequencies, independent control of both the frequency at which optical feedback occurs, and the angle at which radiation is coupled out of the cavity is obtained, without the need to match the effective mode index to a particular value.

The QCL active region employed for our experimental work is a slightly modified version of the 10-µm thick three well resonant phonon active region design, which provides the best QCL temperature performance to date [12]. We engineered our gratings by employing the following refractive index modulation: $n(z) = n_1 + (n_2 - n_1)f\left(z; k_{fb}, k_e(\alpha)\right)$ where $n_1$ and $n_2$ are the minimum and maximum values of the refractive index, respectively, and the function *f*, which varies between 0 and 1, contains periodic functions oscillating at the feedback wavevector $k_{fb}$ and the extraction wavevector $k_e$ (α). The latter is, in turn, a function of the desired beam steering angle α. In the present case, since the grating is patterned on the top metal of the double-metal QCL the only obtainable $f$ is piecewise constant, and $n_2$ and $n_1$ correspond to the effective refractive index of the metallic waveguide with, and without, the top metal, respectively.

The spatial periodicity of the feedback wavevector $k_{fb}$ is tailored on the desired laser frequency $v$, and can be extracted from the formula for first order gratings $k_{fb} = 2n_2 v/c$ where $c$ is the speed of light in vacuum. The latter relation can be assumed to be valid for weak refractive index variation and in the limit of slit widths of a few µm-wide; alternatively, i.e. in the presence of a photonic bandgap, it is a good approximation of the lower edge of the photonic bandgap when $n_1 < n_2$.

The waveguide openings, behaving as localized emitters, ensure that light can be coupled out. The far-field pattern then results from the interference of the radiation emitted by each grating and can be controlled by its periodicity. For emission at a controlled angle α (measured with respect to the waveguide normal (α = 0 for vertical emission)), $k_e$ (α) can be calculated as

$$k_e = \frac{v}{c}(n_2 + sin(\alpha)) \qquad (1)$$

There are several ways to devise a function containing the two spatial harmonics $k_{fb}$ and $k_e$. As a starting point, we developed a set of dual periodicity gratings using a clipped sinusoidal generating function:

$$f(z) = H\big(sin(2\pi k_{fb} z) + sin(2\pi k_e z) + d\big) \qquad (2)$$

where $H(x)$ is the Heaviside step function ($H(x) = 0$ for x<0, $H(x)=1$ otherwise) and $d$ defines the duty cycle of the grating function. Three different gratings were designed, all targeted to a 3.1 THz emission frequency and with extraction angles α = 0° (sample A), α = 30° (sample B), and α = 60° (sample C). The grating parameters for each are summarized in table I, where the design beam steering angles are reported in the $α_1$ column.

From our two-dimensional (2D) simulations, we extracted $n_1$ = 2.55 for the unperturbed waveguide without the top metallization and $n_2$ = 3.48 for the TM mode of the double-metal waveguide. For sample A the design parameters have been tuned to obtain a grating periodicity of 28 µm and a slit width of 4.5 µm, so that the first two grating spatial harmonics correspond to $k_e$ =356 cm$^{-1}$ and $k_{fb}$ = 712 cm$^{-1}$. Using the extracted $n_1$ and $n_2$ values, we performed a plane wave calculation of the photonic bandstructure of the grating, obtaining a photonic bandgap extending from 3.1 to 3.35

THz, well within the QCL gain bandwidth [12], meaning that lasing at both frequencies can in principle occur.

For samples *B* and *C*, the bandstructure cannot be defined since the *n*-modulation is non-periodic; however, we expect the frequency distribution of the resonator modes to exhibit the same energy gap, since this is mostly determined by the grating component that provides optical feedback. For this reason, we used the $k_{fb}$ of *A* and varied $k_e$ according to Eq. (1) to keep the laser frequency unchanged.

Figure 1a shows the spatial dependence of the effective refractive index for samples *A*, *B* and *C*. The $\alpha = 0$ case (sample *A*, black curve) corresponds to a second-order (periodic) grating, while the other gratings are not-periodic. The corresponding Fourier transform of each grating is shown in Fig. 1b. All gratings exhibit a spectral peak at $k_{fb} = 712$ cm$^{-1}$ that provides optical feedback to the laser, and lower frequency peaks for extraction. For the case $\alpha = 0$ (black curve) the extraction peak is at $k_e = k_{fb}/2 = 356$ cm$^{-1}$ (second-order grating) so that a mode propagating in the waveguide is coupled outside in the vertical direction. This process is depicted schematically in the figure by means of a light-cone diagram (dashed circle) drawn with its center corresponding to the value of the guided mode wavevector, and with a radius equal to the free space wavevector at the design frequency (3.1 THz, 103 cm$^{-1}$). The projection of a spectral peak on the circle allows identification of the emission angle, represented by the arrow in the figure. The second and third gratings corresponding to samples *B* and *C*, respectively, exhibit multiple peaks, two of which fall inside the light cone and can couple out radiation at different angles (colored arrows). The additional spectral peaks in the Fig. 1b arise from the harmonic mixing of $k_{fb}$ and $k_e$ through the clipping of the sinusoids in Eq. 2. However, falling outside the light cone, they do not contribute to out-coupled radiation.

The QCL devices were fabricated in a double-metal waveguide configuration [22], taking care of removing the 75-nm-thick n$^+$ contact layer below the grating apertures via reactive ion-etching to reduce the cavity losses. Strong absorbing boundary conditions were implemented to suppress undesired Fabry-Pérot modes, by surrounding the grating pattern with a thin Cr frame (7 nm) extending 20 µm away from each side of it [24].

The emission spectra were acquired using a Fourier-transform infrared spectrometer in rapid-scan mode and equipped with a DTGS detector. Far-field beam profiles were measured at 10 K, with the devices driven in pulsed mode with 200 ns pulses at 50 kHz repetition rate. A two-dimensional mapping of the emission was obtained by scanning a pyroelectric detector over the plane parallel to the device surface at a distance of ~ 6 cm. The acquired data were then mapped onto spherical coordinates centered at the sample, taking into account the detector position with respect to the sphere center. The results are shown in Figs. 2c–h, together with the schematics of the far-field experimental arrangement (Fig. 2a); the measured emission frequencies and angles are summarized in table I. In the experimental configuration of Fig. 2a, α is equal to the far-field elevation ϑ, shown on the vertical axis of Figs. 2f-h. All samples emit single-mode; for samples *A* and *C* the detected emission frequency (Figs. 2c and 2e, respectively) can be assigned to the upper band-edge mode of the calculated photonic bandstructure of sample *A*. However, sample *B* is observed to lase at 3.53 THz (Fig. 1d), a frequency larger than the upper band edge of the calculated photonic bandgap, which we tentatively attribute to emission from a localized mode in the non-periodic grating.

As a common feature, all samples exhibit two lobes in the measured far-field emission pattern (Figs. 2f–h), that in sample *A* (second order grating) (Fig. 2f) suggests that the device is lasing on an antisymmetric mode, as also confirmed by its emission frequency. Samples *B* (Fig. 2g) and *C* (Fig. 2h) both exhibit a pair of lobes, oriented along the $\alpha_m = \pm 18°$ and $\alpha_m = \pm 35°$ directions, where $\alpha_m$ is the measured beam steering angle, here defined to be half the angular separation between the two lobes of the far field pattern. Although this behavior qualitatively matches the predicted larger emission angles for sample *C* with respect to sample *B*, a large discrepancy is found with the designed angles ($\alpha_1$ in Table I). Two effects may explain such discrepancy: i) the observed QCLs emission at a higher frequency requires that the calculated emission angles have to be corrected using such a measured frequency. The predicted angular values consequently shift towards those measured experimentally ($\alpha_2$ in table I); ii) our design methodology based on the light cone diagrams is only applicable for weak refractive index perturbations in periodic structures. Here, the refractive index variation at $k_{fb}$ is sufficiently large to open a 0.25 THz photonic bandgap, and the structure is non-periodic. As such, the model can only be applied qualitatively.

Although the results in Fig. 2 show that it is possible to control the QCL emission angle by tuning the extraction-wavevector of the dual-periodicity grating, it proved difficult to generate controlled single-mode emission at angles larger than 45°. The presence of multiple harmonics in the grating may permit the resonator to operate on complex feedback paths involving different peaks of the grating spectral function, therefore preventing the desired spectral and angular behavior. To address this we fabricated a second set of THz QCLs targeting in-plane emission and employing a different generating function consisting of two superimposed square gratings with different periodicities:

$$f(z) = H\left(S_d(2\pi k_{fb}z) + S_d(2\pi k_e z)\right) \qquad (3)$$

where $S_d(x)$ is a square wave with duty cycle *d*, defined as $S_d(x)=1$ for $-d/2 < x < d/2$ and $S_d(x)=1$ otherwise.

Figures 3a and 3b show the resulting refractive index modulation and its Fourier transform, respectively. Two spectral peaks, at $k_{fb} = 712$ cm$^{-1}$ and $k_e = 462$ cm$^{-1}$ are visible which correspond to the fundamental frequencies of the two superimposed gratings. The intensity of the mixed harmonic peaks due to the clipping effect is here largely reduced compared with that seen previously (Fig. 1a). Figure 3b also shows how the emission angle is different for devices operating on the two band-edge modes. The inner light cone (red dashed line) corresponds to the 3.1 THz mode of the second order DFB (sample *A*), while the outer light cone (green dashed line) corresponds to the 3.35 THz mode. Since the two modes have the same wavevector, the two light cones are centered at the same location, but the difference in their radius causes a difference in the expected extraction angle.

Figures 4a–c show the far-field emission pattern (Figs. 4a and 4b) and the measured laser spectra (Fig. 4c) of two prototype QCL devices (samples *D* and *E*) exploiting a top grating designed via Eq. 3. The far-field measurement has been performed in the configuration depicted in Fig. 2b. In this geometry $\alpha = \vartheta - 90°$ when $\varphi = 0$. The two devices, nominally identical, emit single-mode at two different frequencies due to processing inhomogeneities (3.14 THz, sample D; 3.47 THz sample E), corresponding to the lower and upper bands of the grating. In neither case, in-plane emission is achieved, but THz radiation is emitted along the surface of a cone with the axis oriented along $\theta = 0°$ ($\alpha = 90°$) with divergences of 50° (sample D) and 70° (sample *E*) corresponding to $\alpha_m = 65°$ and $\alpha_m = 55°$ respectively, 17% lower than the corrected nominal angles $\alpha_2 = 80°$ and $\alpha_2 = 66°$.

Figure 4d shows the light-current and voltage-current (LIV) characteristics of sample *E*. Laser action up to 110 K was achieved, corresponding to a lattice temperature of 135 K, [25] i.e. 45 K lower than in the edge-emitting THz QCL fabricated from the same active-region, as expected due to the increased losses induced by the grating and the absorbing chromium edge. Figure 4e shows a comparison of the optical performances of samples *A–D* (gratings *a–d*, respectively). Although no qualitative changes in the LIV curves were observed, the slope efficiency increases with larger extraction angle, whilst the threshold current density ($j_{th}$) remains practically unchanged. The vertical dispersion visible in the slope efficiency data of Fig. 4e is partly due to our uncertainty in the estimation of absolute power (calculated by integrating the far field) and to the multimode and multi-lobed behavior of some devices. The slope efficiency increases as the extracted light becomes progressively more collinear with the lasing waveguide mode, in agreement with what observed in 3$^{rd}$-order and 2$^{nd}$-order DFBs [9,10].

In summary, we have developed THz QCLs operating under the distributed feedback of a dual-periodicity grating whose emission angle and frequency can be lithographically tuned between vertical and 50° angular emission via independent control of the grating periodicity. This opens the way for development of a novel class of in-plane emitting THz DFB QCLs exploiting a different grating concept to achieve single-mode and high-collimated emission with an easier, lithographically independent, phase matching approach.

**Acknowledgements**

This work was partly supported by the Italian Ministry of Education, University, and Research (MIUR) through theprogram FIRB-Futuro in Ricerca 2010, RBFR10LULP, "Fundamental research on terahertz photonic devices," by the ERC advanced grants SoulMan (No. 321122) and Tosca (No. 247375) and by the EPSRC (UK), the Royal Society, and the Wolfson Foundation.

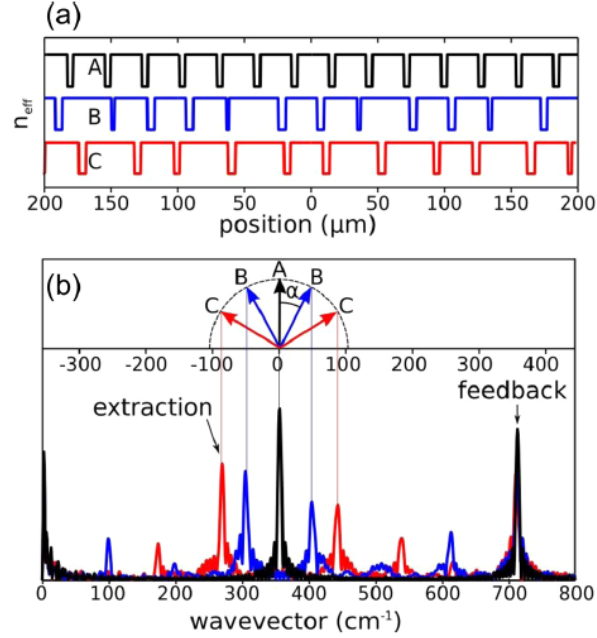

**Figure 1: a)** Spatial dependence of the effective refractive index for samples *A*, *B* and *C*. The refractive index switches between 2.55 and 3.48 in each case. **b)** Spatial spectra of the gratings exploited for samples A, B and C. All gratings have the same feedback peak at 712 cm$^{-1}$. The light cone (dashed circle) has a radius corresponding to the free space wavevector of radiation at 3.1 THz (103 cm$^{-1}$) and is used as a reference to determine the direction of the outcoupled radiation (arrows), based on the location of the extraction peaks.

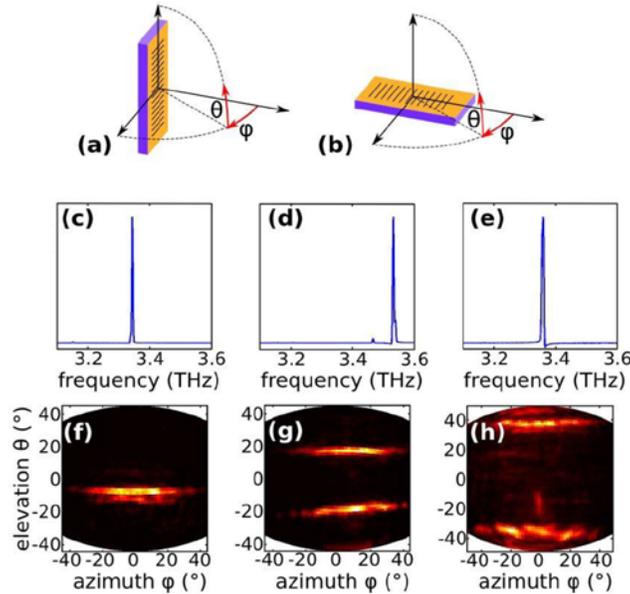

**Figure 2: a,b)** Schematic diagrams showing the orientation of the QCL with respect to the far-field angles in the (a) vertical and (b) edge emission experiments. **c-e)** Emission spectra of (c) sample A, (d) sample B, and (e) sample C. **f-h)** Far-field emission pattern of (f) sample A, (g) sample B, and (h) sample C, measured in the configuration shown in panel (a).

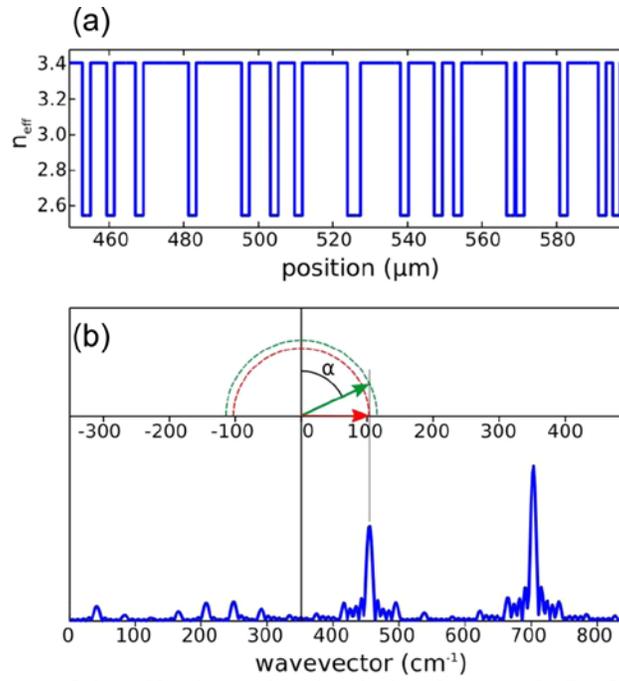

**Figure 3: a)** Spatial dependence of the effective refractive index for sample D. **b)** Spatial spectra of the grating of panel a. Two light cones (dashed lines) are drawn, corresponding to the two bandgap modes of a first order grating with spatial periodicity of 712 cm$^{-1}$. The two modes are predicted to be at 3.1 THz (red, inner circle) and 3.35 THz (green, outer circle), and to radiate at different angles, as indicated by the arrows.

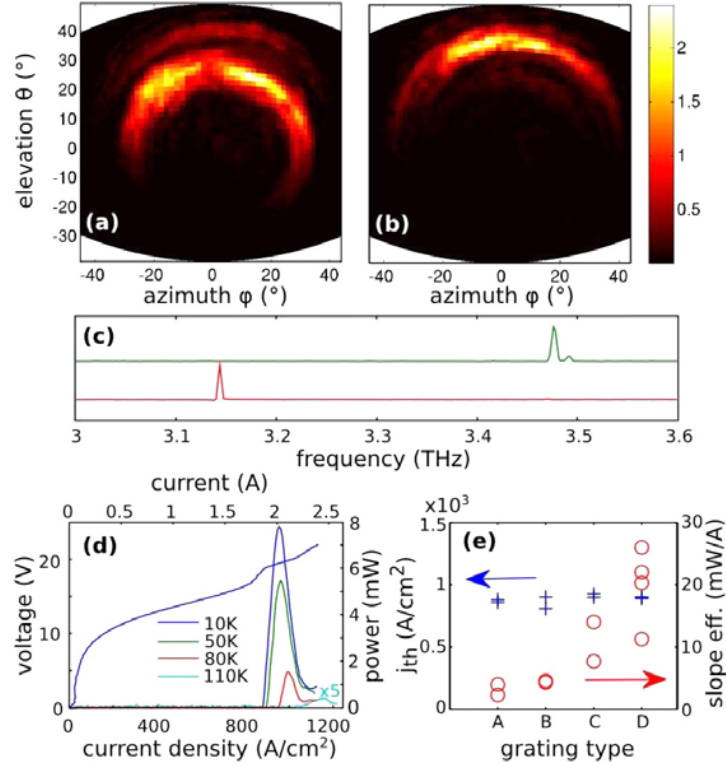

**Figure 4: a,b)** Far field emission pattern of (a) sample *D* and (b) sample *E*, exploiting the same grating but operating on two different bandgap modes. The measurement has been performed with the arragement depicted in Fig. 2b. **c)** Emission spectra of samples *D* and *E*. **d)** Light-current-voltage characteristics of sample *E*. **e)** Comparison of the threshold current density ($j_{th}$) (blue crosses) and slope efficiency (red circles) of a set of samples equipped with different gratings (marked on the x axis).

| Sample | $k_{fb}$ (cm$^{-1}$) | $k_e$ (cm$^{-1}$) | $\alpha_1$ (degrees) | $\alpha_2$ (degrees) | $\alpha_m$ (degrees) | $\nu$ (THz) |
|--------|---------------------|-------------------|---------------------|---------------------|---------------------|-------------|
| A | 712 | 356 | 0 | 0 | 5 | 3.34 |
| B | 712 | 411 | 30 | 26 | 18 | 3.53 |
| C | 712 | 449 | 60 | 53 | 35 | 3.36 |
| D | 712 | 462 | 90 | 80 | 65 | 3.14 |
| E | 712 | 462 | 63 | 66 | 55 | 3.47 |

**Table I**: Design parameters and measured frequency (ν) and steering angles α of THz QCLs exploiting different dual-periodicity gratings. $k_{fb}$ and $k_e$ are the feedback and extraction wavevectors respectively. $\alpha_1$ is the designed beam steering angle at the design frequency of 3.1THz, $\alpha_2$ is the corrected angle taking into account the measured emission frequency, and $\alpha_m$ is the measured steering angle.